# Low-loss terahertz negative curvature suspended-core fiber

Jing Deng, Lei Fan, Qichao Hou, Xingfang Luo, Chun-Fang Rao, Yuan-Feng Zhu

*Abstract*—Inspired by the design concept of negative curvature hollow-core fibers, this paper presents an innovative negative curvature suspended-core THz fiber. Compared to traditional suspended-core fibers, all structural units of this fiber are designed with circular boundaries, effectively avoiding the issues of insufficient mechanical strength and manufacturing difficulties associated with the wide and thin rectangular support arms in traditional structures. The numerical simulation using the full-vector finite element method shows that the optical fiber loss is as low as 0.02cm$^{-1}$ in 0.66-1.09THz, and the low loss bandwidth is 0.43THz. In addition, near-zero flat dispersion of 0.33±0.41 ps/THz/cm can be achieved. The fiber exhibits excellent characteristics of low loss, wide bandwidth, and low dispersion, theoretically opening a new research path for the design of low-loss THz fibers.

## I. Introduction

Terahertz (THz) radiation is an electromagnetic wave ranging from 0.1 THz to 10 THz. This band has attracted a lot of attention from researchers and scientists. It covers a wider range of applications such as medical diagnosis[1], drug testing[2], spectroscopy[3], astronomy[4], safety[5], DNA hybridization[6], non-invasive imaging[7], biosensors[8], agriculture[9], communication[10], etc. Many THz systems still rely on free-space transmission in the air at this stage, however, THz waves are affected by factors such as water vapor when they are transmitted in free-space, resulting in signal loss[11]. Consequently, the design and optimization of low-loss THz wave-guides have become a primary objective for researchers aiming to enhance THz technology. The current THz fiber mainly includes hollow fiber, porous fiber, and suspended core fiber[12-14]. For all kinds of dielectric waveguides, although THz hollow-core fibers exhibit the advantage of low transmission loss[15-18] due to the relatively long wavelength of THz waves, the cross-sectional dimensions must remain relatively large to achieve efficient waveguide transmission[19]. This structural feature significantly reduces flexibility, making them prone to mode leakage or radiation loss when bent, thereby constraining their deployment flexibility in complex environments. Porous core fiber is a method to achieve low loss THz guidance by confining most of the power to the air region. In 2016, Md Saiful Islam et al.[20] proposed a based circular porous fiber, which can be used at an operating frequency of 1THz. They achieved a very low total loss of 0.043cm$^{-1}$, and a super-flat dispersion change of 0.09ps/THz/cm, nevertheless, the structure introduced more air holes in the core part. In 2022, Md Shafiqul Islam et al.[21] proposed a low loss hexagonal porous core THz fiber capable of achieving total loss of 0.0284 cm$^{-1}$ and a dispersion of less than 0.27ps/THz/cm. In 2023, Neel Kumar Arya et al.[22] proposed a low-loss THz fiber whose core part is alternately arranged by circular and ellipse shape air holes. At an operating frequency of 0.9THz, the proposed fiber exhibits a low total loss of 0.0697cm$^{-1}$. However, their fiber all include non-circular structural units in the core, such as ellipse and rectangle air holes, which induces structural complexity. In 2011, Mathieu Roze et al.[23] proposed the THz suspended-core fiber, which is composed of a fiber core, thin support arm, three large air holes. The fiber loss can be less than 0.0868dB/cm (0.02cm$^{-1}$) in the frequency range of 0.28-0.48 THz. The support arm ensures stable connection between the fiber core and the jacketing layer. In addition, the jacketing layer facilitates users to manually operate and control the suspended-core fiber, ensuring that it is not affected by external factors. However, the mere 10μm thickness of its support arm introduces significant challenges in maintaining structural integrity during fabrication. In 2015, Yuan-Feng Zhu et al. [24] a low-loss THz fiber with crossed rectangular dielectric layers. The two rectangular dielectric layers not only serve as supporting arms, but also the area where the two rectangular dielectric layers intersect can serve as the core of the optical fiber. In the frequency range of 0.44 THz to 0.84 THz, the fiber loss can be as low as 0.0868 dB/cm (0.02 cm$^{-1}$) with the dielectric layer thickness of 20μm. and a low dispersion of 0.56-0.98 ps/THz/cm can be achieved. In recent years, other THz suspended-core fiber with different structures have been proposed[25-27], all of which have obtained lower loss and flat dispersion. However, in order to obtain lower losses characteristics, irregular structures such as rectangular, ellipse or semi-ellipse have been introduced into the suspended core, which greatly increases the difficulty of fabrication.

Researchers have also delved into the manufacturing technology of THz suspended core fiber, and have achieved some progress. In Reference [23], researchers used drilling and stacking techniques to achieve the earliest case of THz suspended-core fiber manufacturing. However, due to the thin

This work is supported in part by the National Natural Science Foundation of China (NNSFC) (Grant No. 12064016, 52061017). (Corresponding author: Yuan-Feng Zhu)

Jing Deng, Lei Fan, Xing-fang Luo, Chun-Fang Rao, Yuan-Feng Zhu are with the School of Physics and Communication Electronics, Jiangxi Normal University, Nanchang, Jiangxi Province Nanchang, 330022, P. R. China (e-mail:201926002098@jxnu.edu.cn;fan_lei@jxnu.edu.cn;xfluo@jxnu.edu.cn;rcf0322@jxnu.edu.cn; yuanfengzhu@jxnu.edu.cn).

Qichao Hou is with the Greater Bay Area Institute for Innovation, Hunan University, Guangzhou 511300, China (e-mail: hqc1121@hnu.edu.cn)

and irregularly formed support arms, the fiber experienced deformation of the support structures during fabrication, ultimately significantly constrained the enhancement of THz wave transmission performance. With the development of technology, 3D printing technology is used to manufacture THz suspended-core fibers. As reported in Reference [28], researchers have successfully achieved continuous fabrication of polypropylene-based suspended-core fibers with lengths up to several meters, demonstrating both low transmission loss and near-zero dispersion characteristics. Due to the limited fabrication precision of current 3D printing technology, the fabricated fiber exhibits a core diameter on the millimeter scale, with support arms approximately 0.4 mm thick. Owing to the substantial dimensional scale of the fiber, this fiber is only suitable for applications in the low-frequency band. In addition, limited by the size of the forming cavity, 3D printing technology struggles to achieve the fabrication of ultra-long THz fibers (length > 1 m). Therefore, it is particularly necessary to carry out the design work of a new type of suspended-core THz fiber structure. The designed THz fiber should not only have low-loss characteristics but also possess excellent shape retention ability. Only in this way can the drilling-stacking technology be effectively used to fabricate ultra-long THz fibers.

In this paper, a THz negative curvature suspended-core fiber is proposed. All structural units of the fiber are circular dielectric tubes, which is superior to the typical suspended core fiber structure [29-31]in terms of shape retention during fiber fabrication. Full-vector finite element method (FEM) with PML is used to characterize the waveguide properties of the proposed fiber. The proposed fiber can achieve fiber loss as low as 0.0868dB/cm (0.02cm$^{-1}$) at 0.66-1.09THz, and total loss is only 0.065dB/cm (0.015cm$^{-1}$) at 1.0 THz. Moreover, by adjusting the structural parameters of the fiber, the dispersion of 0.33±0.41 ps/THz/cm can be demonstrated.

## II. FIBER STRUCTURE

The cross-section of the designed THz suspended-core fiber is shown in Fig. 1.(a). The structure of the optical fiber is described by the following parameters: The cladding area is composed of large dielectric tubes with radius $r_1$ and thickness $t$. The number of large dielectric tubes is $N$. The same number of small dielectric tubes are suspended in the core area, with radius $r$ and thickness $t$. To facilitate parameter optimization and performance discussion, we assume to set a small circle with a radius of $r_0$ in the central part. The inner diameter of the dielectric ring $D=2\times[r_0+2\times(r+r_1)]$. Structurally, the composition of this fiber is similar to numerous anti-resonant negative curvature hollow core fibers[32-34] but its optical transmission principle is different from theirs. This is a total internal reflection negative curvature suspended-core fiber, and its transmission principle is the same as that in Reference [35]. Compared with suspended core fibers [23], this fiber adopts a negative curvature structure, which avoids the need for traditional suspended-core fibers to have thin arms to construct the fiber. The fiber core consists of small dielectric tubes and an air area surrounded by them. The electric field diagram of the fundamental mode of the fiber for $N=4$ is shown in Fig. 1.(b). The figure shows that most of the energy of the mode is distributed in the air holes, which effectively reduces the absorption loss of the intrinsic material. At the same time, this mode is strongly confined to the fiber core area. Additionally, compared with traditional suspended-core optical fibers [23], a simple fiber structure with all structural units having circular boundaries will have more prominent advantages in fiber manufacturing [36].

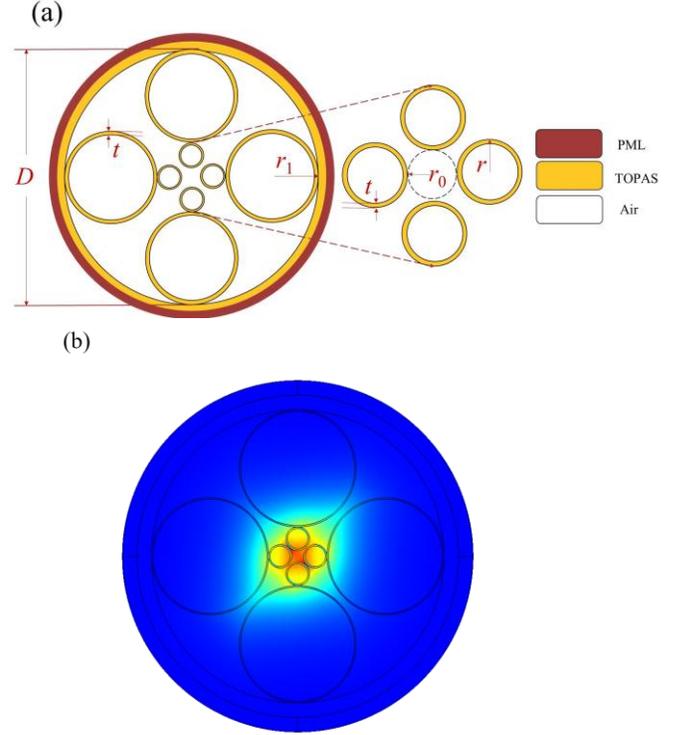

Fig. 1. (a) Cross-section of the proposed THz negative curvature suspended-core fiber, (b) Mode field distribution of fiber at 0.5THz.

A finite element method (FEM) solver (COMSOL) was used to simulate the numerical properties of the fiber under perfectly matched layer boundary conditions. The whole structure contains 151084 domain units and 6736 boundary units. In the air and material medium regions, the maximum element size of the air part is set to $\lambda/4$, and the maximum element size of the medium part is set to $\lambda/60$, in order to ensure the accuracy of the results. All the data used for the numerical study of the fiber structure are calculated. The optical fiber confinement loss is calculated using the following equation[37]:

$$L_c = 8.686\left(\frac{2\pi f}{c}\right)\text{Im}\left(n_{eff}\right)\times 10^{-2} \text{ cm}^{-1} \quad (1)$$

where $L_c$ represents the confinement loss, Im[$n_{eff}$] represents the imaginary part of the effective refractive index, $f$ is the operating frequency, and $c$ is the speed of light.

For a given mode, the material absorption loss is defined by[38]:

$$\alpha_{eff} = \frac{1}{2}\sqrt{\frac{\varepsilon_0}{\mu_0}}\left(\frac{\int_{mat} n_{mat}\alpha_{mat}|E|^2 dA}{\left|\int_{mat} S_z dA\right|}\right)\text{cm}^{-1} \quad (2)$$

where $\varepsilon_0$ and $\mu_0$ are the dielectric electric constant and the permeability of the vacuum, $n_{mat}$ is the refractive index of the material, and $\alpha_{mat}$ is the absorption coefficient of the material. $E$ represents the electric field component, $z$ represents the direction vector, and $S_z$ represents the $z$ component of the Poynting vector.

## III. RESULT DISCUSSION

The following analysis will focus on the loss properties of the fiber structure containing different values of $N$. The operating frequency is set to 0.5THz and the tube thickness is set to $t = 20\mu m$. The change of losses, including confinement loss, absorption loss, and total loss, with $r_1$ is shown in Fig. 2. The figure illustrates that the $r_1$ parameter exhibits a minimal influence on absorption loss, whereas its effect on confinement loss is considerable. It is evident that as $r_1$ increases, the distance between the fiber core and the cladding layer expands, which effectively enhances the fiber mode confinement capability and leads to a gradual reduction of the confinement loss. For $N=3$ and 4, when the $r_1$ value is small, the confinement loss significantly exceeds the absorption loss, causing the total loss to align with the confinement loss curve. As the $r_1$ value increases, the confinement loss decreases to a level much lower than the absorption loss, and the total loss then aligns with the absorption loss curve. It is important to note that the fiber structure proposed in this paper imposes an upper limit on the increase of the $r_1$ value. This is because when $r_1$ reaches a certain value, the dielectric tube units will intersect, and the larger the value of $N$, the smaller the upper limit value of $r_1$. For $N=5$, the upper limit value of $r_1$ that can be taken is not sufficient to reduce the mode confinement loss to a smaller value, so the total loss and absorption loss curves are inconsistent, as shown in Fig. 2.(c). Based on the above analysis, the structural parameter of a large medium tube radius is set as $r_1= 2.0$mm for $N=3$, $r_1 = 1.8$mm for $N=4$, and $r_1=1.2$mm for $N=5$ in the follow-up discussions.

Next, we analyze the effects of the parameters $r$ and $r_0$ on the loss. Fig. 3. (a), (b), and (c) are the false color maps of the loss characteristics for the fiber with $r_0$ in the horizontal axis and $r$ in the vertical axis for $N = 3, 4,$ and 5, respectively. From Fig. 3(a-1), it can be observed that the confinement loss decreases gradually with the increase of $r$. This is because as $r$ increases, the distance between the fiber core and the cladding layer also increases. On the other hand, the confinement loss increases with the increase of $r_0$, as the average refractive index of the fiber core decreases with the increase of $r_0$, leading to a weaker constraint on terahertz waves. As can be seen in Fig. 3.(a-2), the absorption loss increases with the increase of $r$ and decreases with the increase of $r_0$. Meanwhile, as shown in Fig. 3(a-3), the total loss and the absorption loss follow nearly the same change patterns. This phenomenon can be attributed to the fact that the magnitude of the confinement loss is on the order of $10^{-4}$, whereas that of the absorption loss is on the order of $10^{-2}$. As a result, the influence of the confinement loss on the total loss is negligible. In Fig. 3.(b) and (c) series, the change trends of confinement loss are the same as when $N=3$, and there is only a slight difference in absorption loss, which is also due to the different distribution ratios of fundamental mode energy in the material. Fig. 3(c-3) shows the change diagram of the total loss when $N=5$. The difference from the previous one is that the change trend of the total loss is quite different. The reason is that the magnitude of the confinement loss is the same as that of the absorption loss, and the total loss is determined by both the confinement loss and the absorption loss.

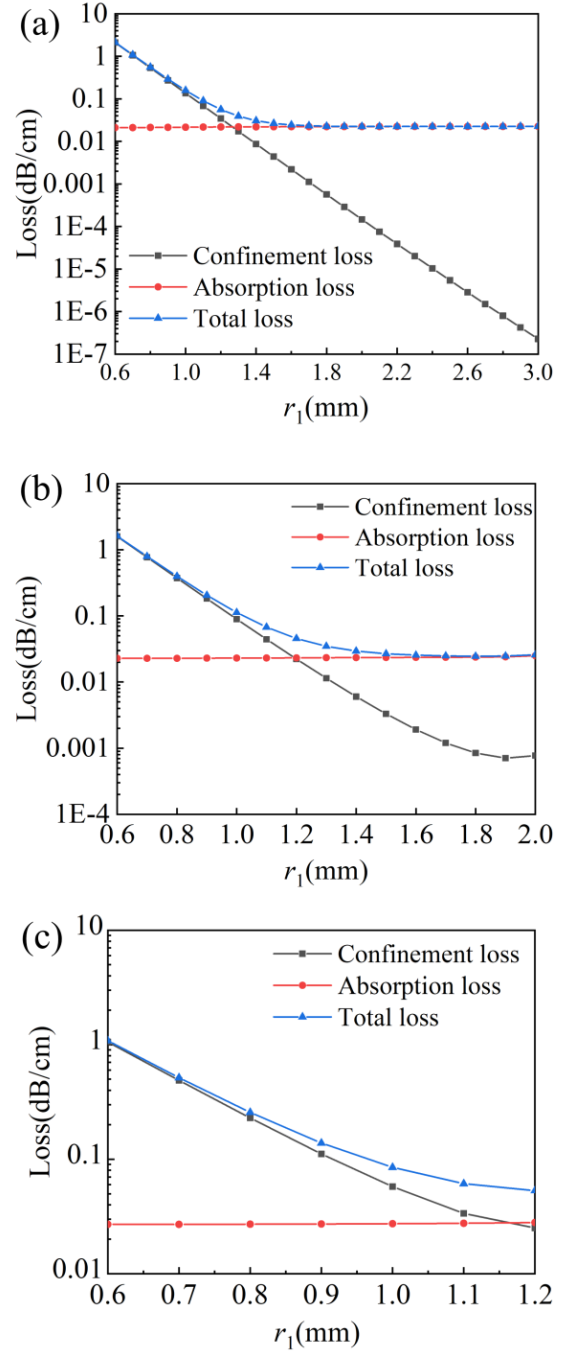

Fig. 2. Loss curves of the fiber with (a) $N= 3$, (b) $N=4$ and (c) $N=5$ versus the $r_1$, other geometrical parameters are set at $r_0$ =300μm, $r$=300μm.

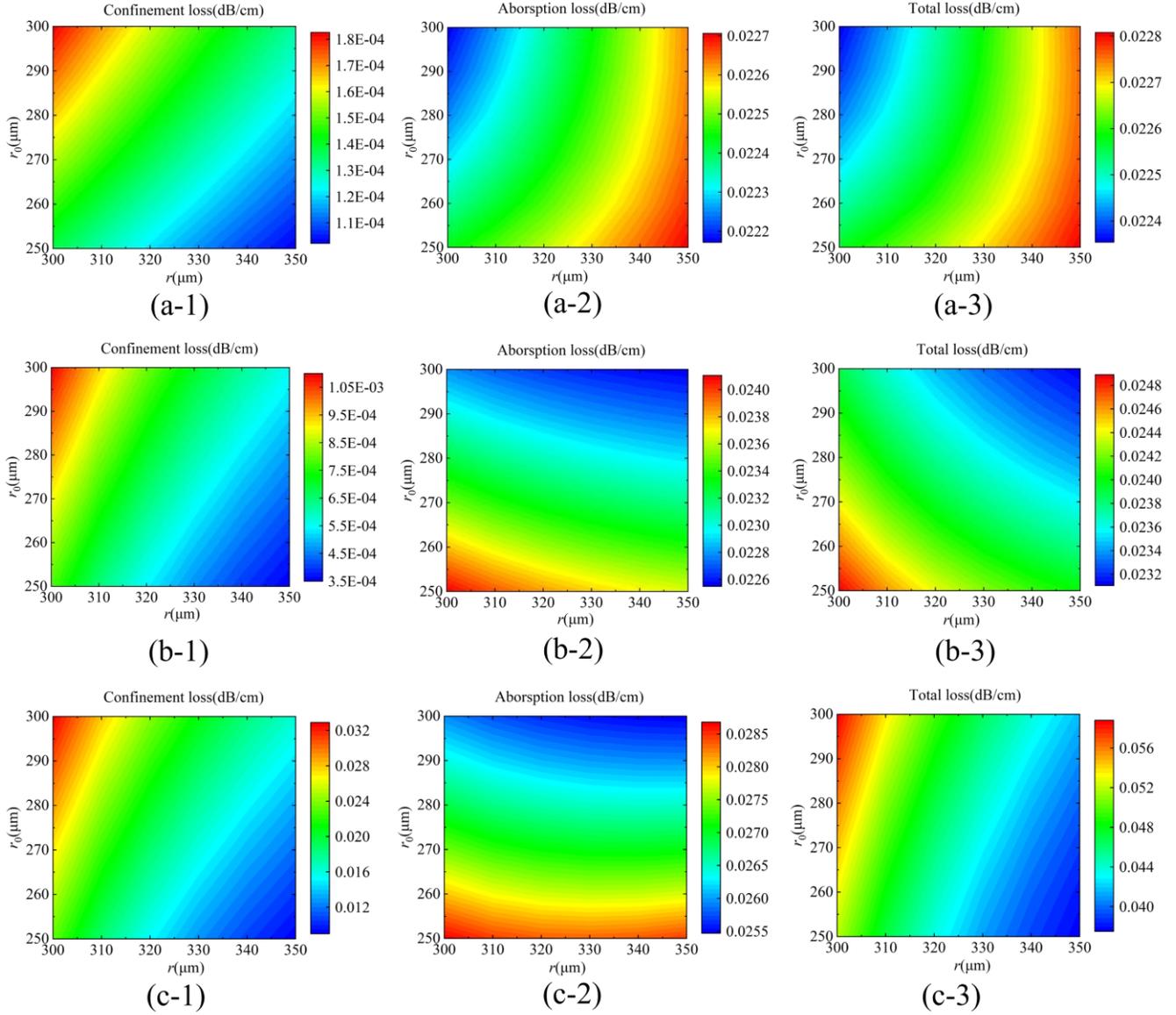

Fig. 3. False color maps of the loss characteristics for the fiber with $r_0$ in the horizontal axis and $r$ in the vertical axis for (a) $N = 3$, (b) $N = 4$ and (c) $N = 5$.

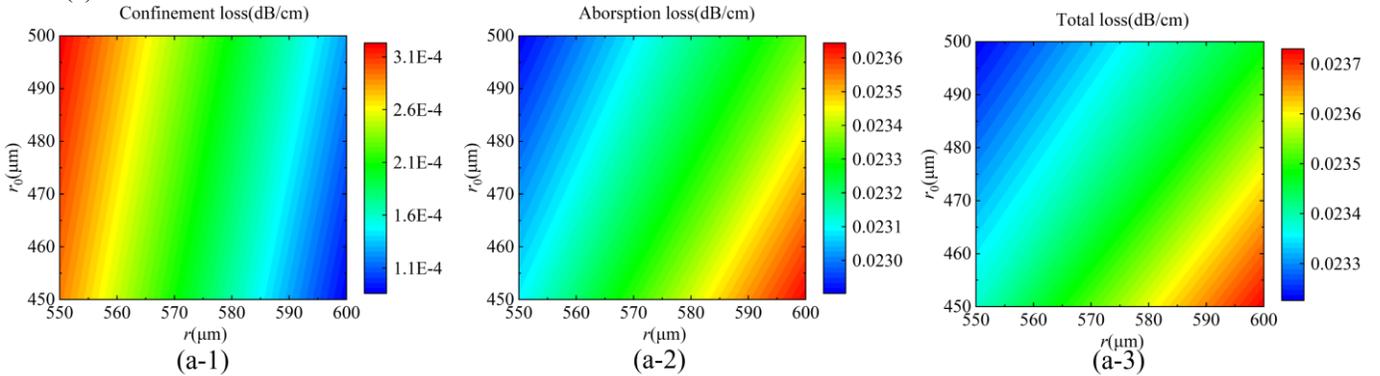

Fig. 4. False color maps of the loss characteristics for the fiber with $r_0$ in the horizontal axis and $r$ in the vertical axis for $r_0$=450μm and $r_1$=1.2mm. and $N = 5$.

In a previous comparative analysis, we observed that when the number of dielectric tubes $N$=5, the loss value increases significantly compared to other configurations, which can be attributed to the fact that the increase in the number of dielectric tubes causes the intersection between the dielectric tubes, which in turn limits the diameter of the dielectric tubes. However, this problem can be solved by adjusting the strategy of the other geometric parameters, as shown in Fig. 4., while the other geometric parameters are $r_0$=450μm and $r_1$=1.2mm. From the results, it can be seen that when $N$=5, adjusting the value of parameters of the fiber can significantly reduce the confinement loss to a value much smaller than the absorption loss, thereby lowering the fiber mode total loss. Overall, whether $N$ is equal to 3, 4, or 5, the fiber can achieve low loss. In other words, this newly designed fiber performs exceptionally well in terms of low loss.

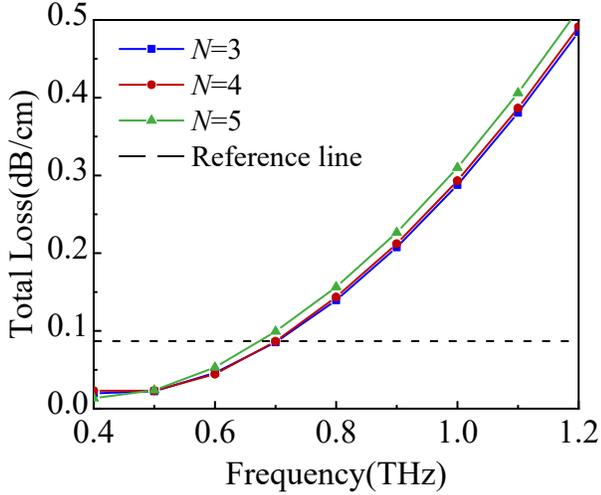

Fig. 5. Total loss as a function of frequency for structures of different dielectric units: $N$=3, $r_0$=300μm, $r$=300μm, $r_1$=2.0mm(blue); $N$=4, $r_0$=300μm, $r$=350μm, $r_1$=1.8mm (red); $N$=5, $r_0$=450μm, $r$=600μm, $r_1$=1.2mm(green).

Fig.5 clearly demonstrates the loss-bandwidth characteristics corresponding to different numbers of dielectric tube units. When the number of dielectric tube units $N$=3 or $N$=4, the loss of the fiber in the frequency range of 0.4-0.7 THz can be stably lower than 0.0868 dB/cm. When $N$=5, the loss of the fiber in the frequency range of 0.4-0.68 THz can also be reduced to 0.0868 dB/cm. In summary, regardless of the variation in the number of dielectric tube units, these fibers exhibit a wide low loss bandwidth characteristic. The following analysis will investigate the impact of other fiber parameters on optical fiber loss characteristics under the condition of $N$=4.

Obviously, reducing the value of $r_1$ can effectively reduce the cross-sectional size of the fiber, thereby enhancing its flexibility. The total loss curves exhibit variations with frequency under different values of $r_1$, as illustrated in Fig. 6.(a). It is observed that as $r_1$ decreases, the ability to confine THz waves weakens, resulting in an increasing of the total loss at low frequencies, but has little effect on the total loss at high frequencies. Furthermore, it is observed that as $r_1$ decreases, the low-loss bandwidth below 0.0868dB/cm contracts significantly, which can be attributed to a substantial increasing of the confinement loss. In subsequent discussions, $r_1$ is set to 1.0 mm, yielding a low-loss range of 0.52-0.7 THz with a bandwidth of 0.18 THz. Fig. 6.(b) and (c) show the total loss as a function of frequency with different $r_0$ and $r$, respectively. Whether the value of $r_0$ or $r_1$ is increased, the duty-cycle of the fiber core will be augmented, consequently leading to an increase in the confinement loss of the fiber mode. This is clearly reflected in the total loss curve at low frequencies. At high frequencies, the dominance of absorption loss over total loss becomes pronounced, rendering the contribution of the confinement loss to total loss negligible. So the two parameters have little effect on the total loss at high frequencies.

With different values of $t$, the total losses as a function of frequency are presented in Fig. 6.(d). with $r_0$=150 μm, $r$=200 μm and $r_1$=1.0 mm. Under the current circumstances, the value presented by the optical fiber diameter $D$ parameter is 5.1 mm. For thickness values of $t$ = 20 μm, 15 μm, and 10 μm, the respective bandwidths maintaining total losses under 0.0868 dB/cm are 0.24 THz (0.51-0.75 THz range), 0.3 THz (0.57-0.87 THz range), and 0.43 THz (0.66-1.09 THz range). To our knowledge, Reference [23] is one of the few examples where suspended-core THz fibers were successfully fabricated using the preform drawing technique. The thin support arm showed significant deformation, which seriously affected the THz transmission characteristics. Therefore, it is necessary to design new THz fiber structures to avoid deformation during the fiber fabrication process. The medium element structure with circular boundaries designed in this article has significant advantages in shape preservation compared to rectangular medium layer elements. And the 0.43 THz bandwidth achieved in this paper for the same $t$=10 μm also demonstrates the potential of the designed fiber to have low loss and wide bandwidth.

For long-distance communication, low dispersion is essential. Because the transmission speed of light at different frequencies is different during long-distance transmission, this leads to the transmission speed of high-frequency light being faster than that of low-frequency light, resulting in changes in the waveform of the light pulse during transmission. Therefore, low dispersion characteristics are also a key focus of THz fiber research. Dispersion includes material dispersion and waveguide dispersion. Here, since the refractive index of the material used in this article remains almost constant within the frequency range studied, only waveguide dispersion is considered, and its calculation formula is as follow[39]:

$$\beta_2 = \frac{2}{c}\frac{dn_{eff}}{d\omega} + \frac{\omega}{c}\frac{d^2 n_{eff}}{d\omega^2}, ps/THz/cm \quad (3)$$

where $\omega$ represents the angular frequency and $n_{eff}$ is the effective refractive index of the fiber mode.

Fig. 7.(a) depicts the variations of the power fractions of the fiber mode as a function of frequency for different values of the parameter $t$. As the frequency increases, the proportion of fundamental mode energy in the material increases, and the modal power gradually transfers from air to the material. Moreover, the thinner the $t$ of the dielectric tube, the lower the proportion of fundamental mode energy in the material, thus achieving lower dispersion values. According to the formula, the group dispersion curve with frequency variation is plotted for different thicknesses, as shown in Fig. 7.(b). When $t$=10 μm, the dispersion value is 0.33±0.41 ps/THz/cm. This low dispersion is beneficial for high bit rates and broadband communication.

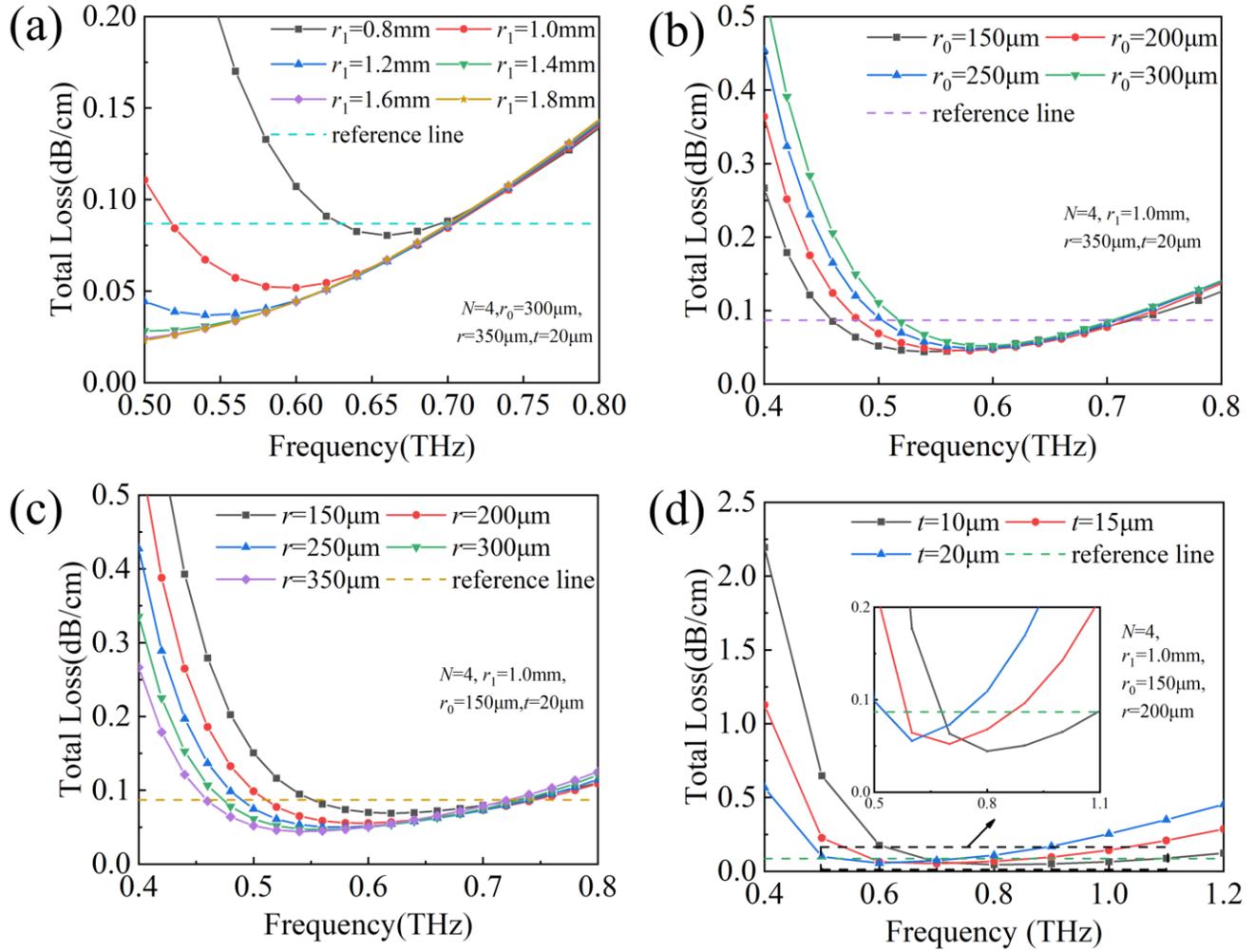

Fig. 6. Fiber loss versus frequency curves for parameters $r_1$ (a), $r_0$ (b), $r$ (c), $t$ (d) at $N = 4$.

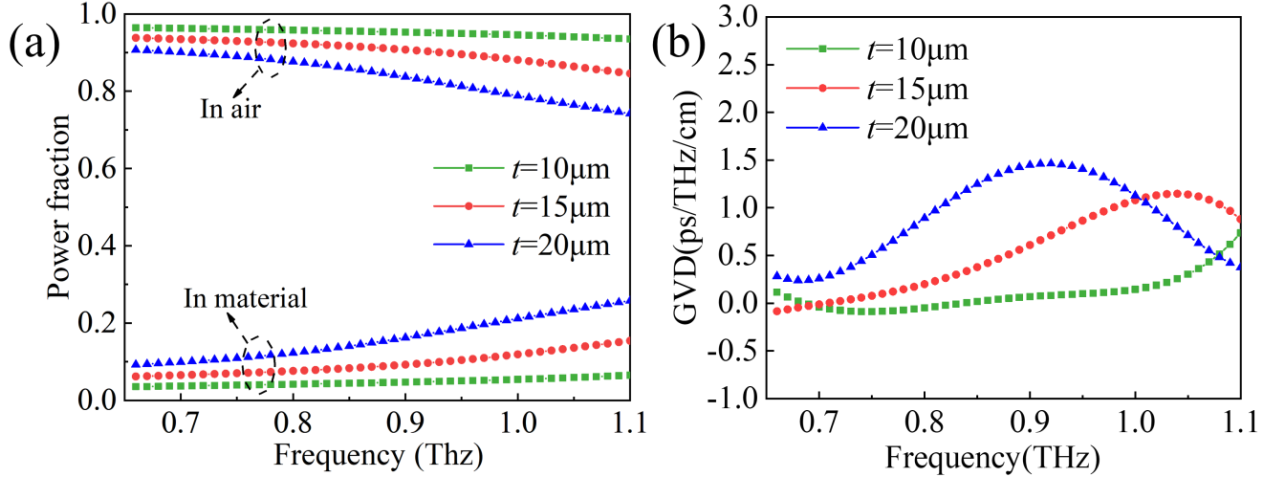

Fig. 7.(a) Power fractions as a function of frequency for different thicknesses, (b) Dispersion as a function of frequency for different thicknesses.

In this study, we conducted a detailed evaluation of the designed THz fiber in key performance indicators such as total loss, bandwidth, and dispersion, and compared and analyzed it with recent research results based on the type of optical fiber. The detailed data is summarized in Table 1. The data in Table 1 shows that at a frequency of 0.4 THz, the total loss of our designed fiber is as low as 0.015 cm$^{-1}$, significantly better than other fibers in the table. In addition, the low loss bandwidth of the fiber reaches 0.43 THz. Compared to it, although the fiber reported in Reference [24] can achieve 0.4 THz at low loss bandwidth, its diameter of 7 mm is much larger than the fiber diameter of 5.1 mm we designed. From Table 1, it can also be observed that porous fibers typically have low loss and dispersion, but their complex shapes, such as containing irregular dielectric elements such as rectangles or ellipses, make the manufacturing process difficult. Although hollow fiber has certain advantages in transmission performance, its large size and resistance to bending limit its applicability in certain application scenarios. In addition, the negative curvature suspended-core fiber in this article exhibits flat and low values in group velocity dispersion (GVD), which helps reduce pulse broadening during signal transmission and improve signal transmission quality.

TABLE 1
COMPARISON OF THE MAIN PROPERTIES OF THE PROPOSED FIBER WITH RECENT RELEVANT REFERENCE

| Reference | Fiber type | Core shape | Total loss ($cm^{-1}$) | Low loss bandwidth (THz) | GVD(ps/THz/cm) |
|---|---|---|---|---|---|
| [20] | Porous | Circle | 0.043(1.0THz) | | 0.09(1.0-1.3T) |
| [21] | Porous | Hexagon | 0.03(1.3THz) | | 0.27(1.0-1.5T) |
| [40] | Porous | Ellipse | 0.045(1.0THz) | | 0.73 ± 0.13 |
| [41] | Porous | Hexagon | 0.1(2.3THz) | | 0.0963(x-polarized) |
| [23] | Suspended | | 0.17 | 0.2 | |
| [24] | Suspended | Rectangle | 0.0092 | 0.4 | 0.56 -0.98(0.44-0.84T) |
| [25] | Suspended | Ellipse | 0.129 | | 0.50±0.35 |
| [26] | Suspended | Rectangle | 0.09 | | 0.03±0.08 |
| Our work | Suspended | Circle | 0.015(1.0THz) | 0.43 | 0.33±0.41 |

## IV. CONCLUSION

In conclusion, a THz negative curvature suspended-core fiber with low transmission loss is proposed. Because the whole structure is composed of circular boundaries, the structure is simple and easy to fabricate. Flexible and adjustable performance can be achieved including transmission loss, dispersion and bandwidth, for different application scenarios by changing important parameters, the results confirm that the proposed fiber can achieve loss as low as 0.0868dB/cm with a low loss band width of 0.43THz. In addition, near-zero flat dispersion of 0.33±0.41 ps/THz/cm can be obtained.